%% file: fest_Scott_TR.tex
\documentclass[runningheads]{llncs}   	
\usepackage{amsmath,amssymb}

\usepackage{tikz}
\usetikzlibrary{decorations.markings,calc}
\usetikzlibrary{shapes.geometric,positioning}
\usepackage{tikz-3dplot}
\newtheorem{thm}{Theorem}
\newtheorem{defn}{Definition}
\newtheorem{prp}{Proposition}

\newtheorem{thesis}{Thesis}

\newcommand{\BS}{{\mathbf B}} % Base space
\newcommand{\HS}{{\mathbf H }} % Fiber space
\newcommand{\FBS}{{\mathbf G }} % Total space
\newcommand{\BP}{{\mathbf {P_{B} }}} % Base point
\renewcommand{\paragraph}[2]{\subsubsection #1{ #2}}
\newcommand{\TM}{{\sl TM}} % Turing machine
\newcommand{\TMs}{{\sl TMs}} % Turing machines
\newcommand{\PTM}{{\sl PTM}} % Persistent Turing machine
\newcommand{\PTMs}{{\sl PTMs}} % Persistent Turing machines
\newcommand{\environment}{{\sl environment}} % Persistent Turing machines

\newcommand{\ITS}{{\sl ITS}} % Interactive Transition system
\newcommand{\ToM}{{\sl TTM}} % Persistent Turing machine
\newcommand{\C}{{\mathcal{P}}} % configuration 
\newcommand{\E}{{\mathcal{O}}} %environment
\newcommand{\B}{{\mathcal{B}}} %behaviour
\newcommand{\MM}{{\mathcal{M}}} %machine
\newcommand{\TS}{{\mathcal{S}}} % topological space
\newcommand{\SC}{{\mathcal{S_{P}}}} % simplicial complex
\newcommand{\Q}{{\mathcal{Q}}} % quiver
\newcommand{\G}{{\mathcal{G}}} % gauge group
\newcommand{\La}{{\mathcal{L}}} % language
\newcommand{\A}{{\mathcal{A}}} % stream algebra
\newcommand{\bb}{{\beta_{\E}}} % domain associated to an environment
\newcommand{\M}{{\mathbb{M} }} % class of PTMs 
\newcommand{\T}{{\mathbb{T}}} % class of ITSs 
\newcommand{\LL}{{\textsf{L} }} % Left move
\newcommand{\R}{{\textsf{R} }} % Rightmove
\newcommand{\m}{{\textsf{m}}} % mapping = transition relation

\begin{document}
\title{Topological Interpretation of Interactive Computation\\
}
\titlerunning{Topological Interpretation of Interactive Computation}
\author{Emanuela Merelli\inst{1}
\and Anita Wasilewska \inst{2}}

\authorrunning{E. Merelli, A. Wasilewska}
\institute{ Department of Computer Science, University of Camerino, Italy\\
\email{emanuela.merelli@unicam.it}\\[.15cm]
\and Department of Computer Science, Stony Brook University,
Stony Brook, NY, USA \\
\email{anita@cs.stonybrook.edu}
}

\maketitle
\begin{abstract}
It is a great pleasure to write this tribute in  honor  of Scott A. Smolka on his 65th birthday. We revisit  Goldin, Smolka hypothesis  that  \textit{persistent Turing machine} (PTM) can capture the intuitive notion of  \textit{sequential interaction computation}.  We  propose  a topological setting to model the abstract concept of  \textit{environment}. We use it to define  a notion of a \textit{topological Turing machine} (TTM) as a universal model for interactive computation and possible model for concurrent computation.


\end {abstract}
\section{Introduction}
\label{sec:introduction}
In 2004, Scott A. Smolka worked with Dina Goldin~\footnote{The work was developed in connection of the celebration of Paris Kanellakis for his 50th birthday. They were his first and last Ph.D student.} and colleagues on a formal framework for interactive computing; the {\sl persistent Turing machine (PTM)} was at the heart of their formalization~\cite{Goldin2004,Goldin2000,Goldin2006}. A \PTM\ is a {\sl Turing machine} (\TM) dealing with {\sl persistent sequential interactive computation} a class of computations that are sequences (possibly infinite) of non-deterministic {\sl 3-tape} \TMs. A computation is called sequential interactive computation because it continuously interacts with its {\sl environment} by alternately accepting an input string on the {\sl input-tape} and computing on the  {\sl work-tape} a corresponding output string to be delivered on the {\sl output-tape}.  The computation is {\sl persistent}, meaning that the content of the work-tape {\sl persists} from one computation step to the next by ensuring a {\sl memory} function. \\
The definition of \PTM\ was based on Peter Wegner's  {\sl interaction theory} developed to embody distributed network programming.

\begin{center}{\sl \small Interaction is more powerful than rule-based algorithms for computer problem solving, overturning the prevailing view that all computing is expressible as algorithms~\cite{Wegner1997,Wegner1998}.}\end{center}
Since in this framework interactions are more powerful than rules-based algorithms they are not expressible by an initial state described in a {\sl finite terms}. Therefore,  one of the four Robin Gandy's principles (or constraints) for computability is violated, as stated in ~\cite{Gandy1980}. 
The need to relax such constraints  allows one to think that interactive systems might have a richer behavior than algorithms, or that  algorithms should be seen from a different perspective. Although \PTM\ makes the first effort to build a \TM\ that accepts {\sl infinite input}, we strongly support the idea that the interaction model should also include the formal characterization of the notion of \environment. 

In this paper, we focus on Smolka et al. original point of view on {\sl persistent} and {\sl interactive computation}. We revisit and formalize a concept of {\sl computational environment} for \PTM\  following Avi Wigderson's machine learning paradigm in~\cite{Wigderson2018}.

\begin{center}{\sl Many new algorithms simply $'$create themselves$'$ with relatively little intervention from humans, mainly through interaction with massive data\footnote{https://www.ias.edu/ideas/mathematics-and-computation}.} \end{center} 
We use the notion of {\sl computational environment} to define class of abstract computable functions as sets of relations between inputs and outputs of \PTM. The computational environment depends on {\sl time} and {\sl space}. It can evolve and so the effectiveness of these functions depends on a given moment and a given context.  

Computational environment is defined in terms of {\sl ambient space}. The ambient space is a generalization of a notion of ambient manifold introduced in~\cite{Garrone2006} to describe the topological quantum computation model. 

\begin{figure}[th!]
\begin{center}
\includegraphics[width=12cm]{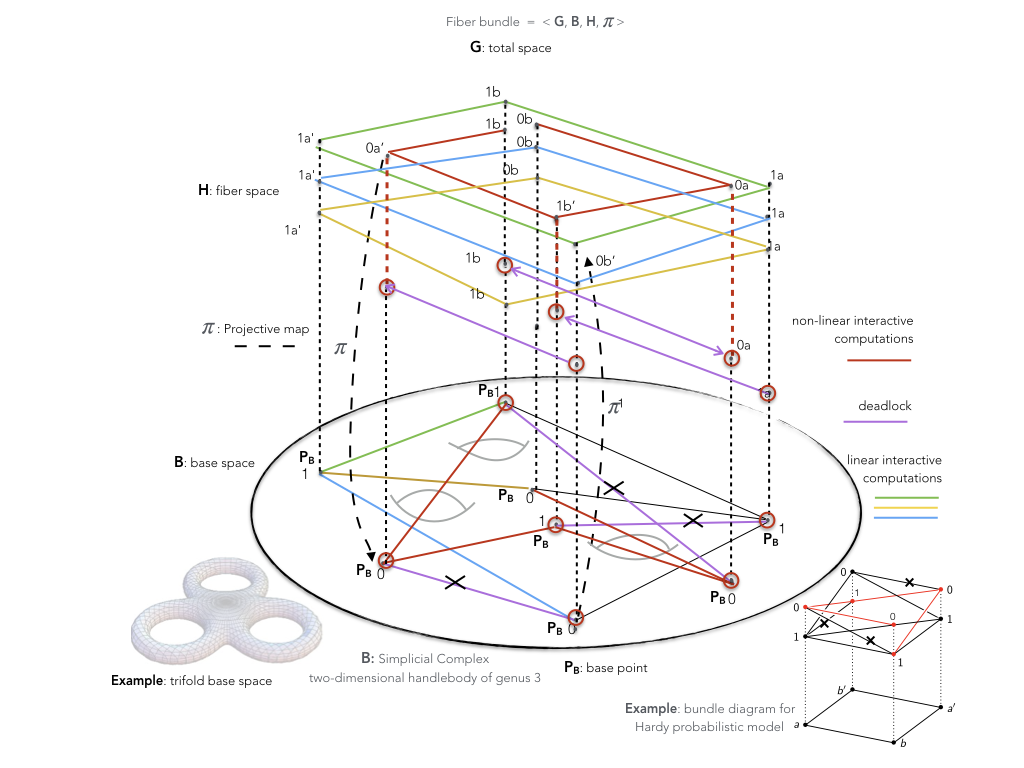}
\caption{Example of topological interpretation of computation. The {\sl base space } $\BS$ is a two-dimensional handlebody of genus 3, such as a trifold.  The small {\sl red} circles around some points of the {\sl fiber space} $\HS$ indicate the presence of states that make the computation inconsistent. The {\sl violet} lines over the {\sl base space} $\BS$ show the corresponding unfeasible paths to be avoided due to the topological constraints imposed by the base space. The non-linear transformations of the fibers states, induced by the {\sl projection map} $\pi$ over the simplicial complex, guarantees the choice of the admissible paths with respect to the topology of the base space. 
The lines marked with a black cross correspond to inconsistent states of the system, which do not exist in the topological interpretation.  The picture at the bottom right corner is an example of computation, it refers to the notion of contextuality~\cite{Abramsky2014b}, informally a family of data -- a piece of information -- which is locally consistent and globally inconsistent.}\label{img:topo}
\end{center}
\end{figure}

We do it in such a way that the infinite computation can be reduced to a set of {\sl relations}, constrained within its  ambient space by loops of {\sl non-linear interactions}. The ambient space is not necessarily a vector space, hence there is a problem of linearity and non-linearity of computation. The non-linearity originated from the {\sl shape} that can be associated to the ambient space, which can be obtained by the topological analysis of the set of {\sl data} provided by the {\sl real} environment. Figure~\ref{img:topo} shows the synthesis of this concept.
The ambient space and \PTM\ can be thought as mathematical representation of complex systems, merely defined as systems composed of many non-identical elements, constituent agents living in an environment and entangled in loops of non-linear interactions. 

We built a topological \PTM\ to model both the behavior of an interactive machine and its computational environment. The main idea of the generalization is that output-tape is forced to be connected to the input-tape through a {\sl feedback loop}. 
The latter can be modeled in a way that the input string can be affected by the last output strings, and by the current {\sl state} of the computational environment. 
A state of a topological \PTM\ becomes a set of input and output relations constrained to an environment whose geometric representation formally defines the {\sl context} of the computation. 
If many topological $PTM$s share the same computational environment,  the computation becomes a stream of interactions of {\sl concurrent processes}, which at higher dimension can be seen as a collection of streams, such as an $n$-string braid as examined in topological quantum information~\cite{Garrone2006}. In this scenario, the computational environment, envisaged as a {\sl discrete} geometric space, may even {\sl evolve} while computations take place. 

The informal description given above depicts the environment. We define it as follows. Given a \PTM, let $X$ be a set of its input and output strings. Since the computational environment depends on {\sl time} and {\sl space}.  In this case the time is represented by collection of steps. For each step $i$ in time, we define an equivalence relation   $\sim_{i}$ on $X$ such that $input_{i}$ in $X$ there exists an operator $f_{i}$ such that $f_{i}(input_{i}) = output_{i}$.

In classical Turing machine the set of operators $f_{i}$ is called rules or transformations. Our goal is to build an environment where this set of functions $f_{i}$ can be discovered. Each element of $X$ represents a transition from one state of the machine to a next guided by the operator $f_{i}$ (unknown for the model) constrained over the computational environment.  The mathematical objects we are looking for should reflect the collective properties of the set $X$ in a natural way to support the discovery of the set of operators $f_{i}$.  These operators allow us to represent $X$ as a union of quotient spaces of the set of equivalence classes $X/\sim_{i}$ of all the {\sl feasible} relations hidden in $X$. The resulting functional matrix of $f_{i}$, also called {\sl interaction matrix}, represents the computational model or what we called the learnt algorithm~\cite{Merelli2015}. 

In order to characterize the set of  operators $\{f_{i}\}$, we decided to analyze the set $X$ of environmental data by a {\sl persistent homology}, a procedure used in topological data analysis (TDA). TDA is a subarea of the computational topology that filters the optimal space among {\sl simplicial complexes}~\cite{Carlsson2009}. A simplicial complex can be seen as a generalization of the notion of graph, where the relations are not binary between vertices, but n-ary among simplices. A simplex expresses any relation among points. For example, a $0$-dimensional simplex is a unary relation of a single point, a $1$-dimensional simplex is a binary relation of two points (a line),  a $2$-dimensional simplex is a three points relation (a full triangle), and so on. For the interested reader, Appendix~1 gives some useful definitions for algebraic and computational topology. Although a simplicial complex allows us to shape the environment as a discrete topological space, the new model of \PTM\  also requires to express the feedback loop between the output at step $i$ with the input at the next step $i+1$ of the computation. To this end, we follows a recent approach proposed in the context of big data and complex systems for embedding a set of correlation functions (e.g. the encoding of a given data set) into a {\sl field theory of data} which is relied on a topological space naturally identified with a simplicial complex~\cite{Rasetti2016}. The resulting mathematical structure is a {\sl fiber bundle}~\cite{Steenrod1951}, whose components are summarized in Figure~\ref{img:topo}.

\noindent The framework consists of three topological spaces $\BS, \HS, \FBS$, and one projection map $\pi$. The {\sl base space} $\BS$, the set of input/output strings embedded in a simplicial complex; the {\sl fiber} $\HS$, the set of all possible computations (the set of $f_{i}$) constrained by the $'$gauged$'$ transformations over the base point $\BP$ of the fiber;  the {\sl total space} $\FBS$ of the fiber bundle obtained by the product of the other two spaces ($\FBS=\BS\times\HS$), and {\sl the projection map} $\pi: \FBS \rightarrow \FBS/\HS$ that allows us to go from the total space $\FBS$ to the base space $\BS$ obtained as a quotient space of the fiber $\HS$. In Figure~\ref{img:topo}, the $\pi$ projection map is represented by dashed lines and used to discover if the geometry of the  base space can constrain the ongoing computation in order to predict and avoid unfeasible transformations, the red lines in the figure. 
In our model, the obstructions that characterize the ambient space and constraint the computation are represented by the presence of n-dimensional holes ($n>1$) in the geometry of the topological space. In our framework the holes represent the lack of specific relations among input and output of topological Turing machine.
It means that the topological space, in our representation as simplicial complex,  has a non trivial topology. As an example, in Figure~\ref{img:topo}, the base space $\BS$ is two-dimensional handlebody of {\sl genus} 3. 
The formal description of the proposed approach rests on three pillars: i) {\sl algebraic and computational topology} for modeling the environment as a simplicial space $\BS$; ii) {\sl field theory} to represent the total space $\FBS$ of the machine as a system of global coordinates that changes according to the position $\BP$ of the observer respect to the reference space $\HS$, and  iii){ \sl formal languages} to enforce the semantic interpretation of the system behaviour into a logical space of geometric forms, in terms of operators $f_{i}$ that here we call correlation functions in the space of the fiber $\HS$.\\  
Consequently, an {\sl effective} \PTM\ is nothing but a change of coordinates, consistently performed at each location according to the  $'$field action$'$ representing the {\sl language} recognized by the machine. \\
%he toolkit to implement the (semantic) representation of the transformations presiding the field evolution. This is the way to study the syntactical aspects of languages generated by the field theory through its algebraic structure. 
While the algorithmic aspect of a computation expresses the effectiveness of the computation, the topological field theory constraints the effectiveness of a computation to a specific environment where the computation might take place at a certain time in space. 

%The {\sl interaction} between the is captured as the reachable subalgebra (subgroup) generated in the semi-direct product of two algebras, the one for the computation and the other for the ambient space.
%
It is right here to recall Landin's metaphor of the ball and the plane, introduced to describe the existence of a double link between a program and machine~\cite{Landin1970}:

\begin{center}{\sl \small One can think of the ball as a program and the plane as the machine on which it runs. ... the situation is really quite symmetric; each constrains the other~\cite{Abramsky2014a}.}
 \end{center}

%Interaction machines, which extend the Chomsky hierarchy, are modelled by interaction grammars, and are able to capture fuzzy concepts like open systems and empirical (data-based) computer science. 

Alan Turing himself, in his address to the London Mathematical Society in 1947, said 
\begin{center}{\sl \dots if a machine is expected to be infallible, it cannot also be intelligent}~\cite{Turing1947}.\end{center} 
It is  becoming general thinking that intelligence benefits from {\sl interaction} and evolves with  something similar to {\sl adaptability checking}~\cite{Merelli2015}. Accordingly, the \PTM, and its topological interpretation seem to be a good starting point for modeling  {\sl concurrent processes} as interactive \TM s~\cite{Abramsky2006}. Also considering that the set of $PTM$s reveals to be isomorphic to a general class of effective transition systems as proved in Smolka et al. in~\cite{Goldin2004}. This result allows to make the hypothesis that the \PTM\ captures the intuitive notion of sequential interactive computation~\cite{Goldin2000}, in analogy to the Church-Turing hypothesis that relates Turing machines to algorithmic computation.
%
%%%%%%%%%%%%%%%%%%%
%
%\input{theory_comp}
%
%\label{sec:theory-computation}
\paragraph*{What is computation?}
\label{sec:computation}
Turing, Church, and Kleene independently formalized the notion of computability with the notion of Turing machine, $\lambda$-calculus, partial recursive functions. Turing machine manipulates strings over a finite alphabet, $\lambda$-calculus manipulate $\lambda$-terms, and $\mu$-recursive functions manipulate natural numbers. The Church-Turing thesis states that 

\begin{center}{\sl every effective computation can be carried out by a Turing machine} or equivalently {\sl a certain informal concept (algorithm) corresponds to a certain mathematical object (Turing machine)}~\cite{Papadimitriou1998}.\end{center}
The demonstration lies on the fact that the three notions of computability are formally equivalent. In particular, the Turing machine is a model of computation like a finite states control unit with an unbounded tape used to memorize strings of symbols. A deterministic sequence of computational steps transforms a finite input string in the output string. For each step of the computation, a Turing machine contains all the {\sl information} for processing input in output, an algorithmic way to computing a function, those functions that are {\sl effectively computable}.  
The $Universal$ \TM\ is the basic model of all effectively computable functions,  formally defined by a mathematical description. 
%\pagebreak
 \begin{defn}[Turing machine] A Turing machine (\TM) is $\mathcal{M}=\langle Q, \Sigma, \mathcal{P}\rangle$,
 \begin{itemize}
\item  $Q$ is a finite set of \emph{states}; 
\item $\Sigma$ is a finite \emph{alphabet} containing \emph{the blank symbol} $\#$; \LL\ and \R\ are special symbols.
\item $\mathcal{P} \subseteq Q \times \Sigma \times \Sigma \times Q \times \{\LL,\R\}$, the set of configurations of $\mathcal{M}$.
\end{itemize}
A \emph{computation} is a chain of elements of $\mathcal{P}$ such that the last one cannot be linked to any possible configuration of $\mathcal{P}$.
 \end{defn} \label{def:TM}
 
 \medskip
 \noindent The {\sl multi-tape Turing machine} is a \TM\ equipped with an arbitrary number $k$ of tapes and corresponding heads.
 
 \begin{defn}[k-tape Turing machine] A non-deterministic k-tape \TM\ is a quadruple $\langle Q, \Sigma, \C, s_{0} \rangle$, 
where 
\begin{itemize}
\item  $Q$ is a finite set of \emph{states}; $s_{0}\in Q$ is the \emph{initial state} and $\textsf{h}\notin Q$ is the\ \emph{halting state}.
\item $\Sigma$ is a finite \emph{alphabet} containing \emph{the blank symbol} $\#$. \LL\ and \R\ are special symbols.
\item $\C \subseteq 􏰆Q \times \Sigma^{k} \times (Q \cup \{\textsf{h}\}) \times  (􏰅\Sigma \cup\{\LL,\R\}) ^{k}$ is the set of configurations.
\end{itemize}
\end{defn}

\noindent The machine makes a transition from its current configuration (state) to a new one (possibly the halt state $\textsf{h}$). For each of the $k$ tapes, either a new symbol is written at the current head position or the position of the head is shifted by one location to the left ($\LL$) or right ($\R$). 

The above definitions of \TM s do not take into account the notion of \environment; the input is implicitly represented in the  configurations $\mathcal{P}$ of $\mathcal{M}$ machine modulo {\sl feasible relations}. The objective of this contribution is to represent the \environment\ explicitly in a way such that the {\sl admissible relations} are naturally determined. Our view is supported by a recent, even though not formal, definition of computation.
\begin{center}{\sl \small Computation is the evolution process of some \environment\ via a sequence of simple and local~steps~\cite{Wigderson2018}.} \end{center}

\paragraph*{A Computational environment} is the base space over which the process of transformation of an input string happens. For the \TM, an \environment\ is any configuration of $\mathcal{P}$ of a machine $\mathcal{M}$, from the initial one to the final one. It is a closed set  -- represented by the functional matrix,  -- whose feasible relations should be known a priori to assure the algorithmic aspect of the computation. Indeed, in \TM\ the \environment\ does not evolve, it remains unchanged during the computation. 
%The environment is reduced to a set of symbols (memorised on the tape),  one of which, at each step, represents the {\sl input} for the control that may change the configuration of the tape, while the
%\bigskip

If we consider the \environment\ as an open set - the set of configurations may changes along the way due to computation - accordingly, the set of feasible relations may change. As Section~\ref{sec-topo} describes, one way to capture this variation is to associate a topology to the space of all possible configurations and use the global invariants of the space to classify the relations in categories whose elements are isomorphic to those of some model of computation, such as the \TM.  In this setting, the local steps (feasible relations) -- the functional matrix -- are affected by global topology. As a consequence, the evolution of an \environment\ corresponds to a change of the topological invariants. Then the classical \TM\ is equivalent to working with a space of states whose topology is invariant, which allows the process of transformation to run linearly. 

While an {\sl interactive computation} takes into account the {\sl non-linearity} of the computation due to the structure of the transformations characterizing it. The non-linearity is implied by the topology of the base space $\BS$, and induced by the semi-direct product factorization of the transformation group, the simplicial analog of the mapping class group, denoted by $\mathcal{G}_{\mathcal{MC}}$. 
%$\mathcal{G}_{A}\wedge$

In the viewpoint of computation as a process, the global context induces {\sl non-linear interactions} among the processes affecting the semantic domain of the computation. The semantic object associated to \TM, that is the function that \TM\ computes, or the formal language that it accepts, becomes an {\sl interactive transition system} for a \PTM. In the topological setting it changed into the pair of $\langle$function, structure$\rangle$, entangled as a unique object. The function represents the behavior and the structure the context. Formally represented by the fiber subgroup in the semi-direct product form of the group of computations (connected to process algebra), denoted by $\mathcal{G}_{\mathcal{AC}}$, and $\mathcal{G}_{\mathcal{MC}}$ the group of self-mapping of the topological spaces (the \environment  self-transformations algebra, i.e. automorphisms which leave the topology invariant), quotient by the set of feasible relations. The new semantic object, a {\sl gauge group} $G=\mathcal{G}_{\mathcal{AC}} \wedge \mathcal{G}_{\mathcal{MC}}$, provides another way to understand the meaning of $contextuality$~\cite{Abramsky2011}, as a tool to distinguish effective computation from interactive computations. That is to identify configurations that are $'$locally consistent, but globally inconsistent$'$, as shown in Figure~\ref{img:topo} and informally summarised in the following sentence 

\begin{center}{\sl Contextuality arises where we have a family of data which is locally consistent but globally inconsistent.} \end{center}

% Hardy model
%we have two experimenters Alice and Bob who can choose between two dichotomic measurements each (a1,a2 for Alice and b1,b2 for Bob). Thus, we have X = {a1,a2,b1,b2}, M = {{a1,b1},{a1,b2},{a2,b1},{a2,b2}} and Om = {0,1} for all m ∈ X. Notice that this is a Bell-type scenario. The rows of the table correspond to the contexts, and the events marked with ’1’ are the ones deemed possible by the model. At the base of the diagram lies the measurement complex. Above each vertex is a fiber representing the two possible outcomes for the corresponding measurement. Possible sections of the empirical model are represented by edges connecting points in the fiber above each context. Global sections correspond to a choice of one section per context such that they all agree at intersections, and appear as closed loops around the bundle. For instance, we have highlighted a global section in blue.

Section~\ref{sec-topo} introduces the new interpretation. {We leave the formal definition and full formalization of the theory corresponding to the group of computations for an evolving \environment\ as future work.}

%%%%%%%%%%%%%%%%%%%
\section{Interactive Computation }
\label{sec:IC}
n this section, we recall the definition of the persistent Turing Machine, \PTM\ as defined by Smolka et al. in~\cite{Goldin2004} and the related notion of \environment\ introduced in their earlier work~\cite{Goldin2000}. We introduce the definitions needed to support the construction of a new topological model that is a generalization of the \PTM. The new model allows one to re-interpret the classic scheme of computability, which envisages a unique and complete space of problems.

The \PTM\  provides a new way of interpreting \TM\ computation, based on dynamic stream semantics (comparable to behavior as a linear system). A \PTM\ is a non-deterministic 3-tape \TM\ (N3TM) that performs an infinite sequence of classical \TM\  computations. Each such computation starts when the \PTM\ reads input from its {\sl input-tape} and ends when the \PTM\ produces an output on its {\sl output-tape}. The additional {\sl work-tape} retains its content from one computational step to the next to carry out the persistence.

\begin{defn}[Smolka, Goldin  Persistent Turing machine] A persistent Turing machine (\PTM) is a ${N3TM}$ having a read-only \emph{input}-tape, a read{\scriptsize{/}}write \emph{work}-tape, and a write-only \emph{output}-tape. 
\end{defn}

\noindent Let $w_{i}$ and $w_{0}$ denote the content of the input and output tapes, respectively, while $w$ and $w'$ the content of work-tape, and $\#$ empty content, then 
\begin{itemize}
\item an {\sl interaction stream} is an infinite sequence of pairs of $(w_{i},w_{o})$ representing a computation that transforms $w_{i}$ in $w_{o}$;

\item a {\sl macrostep} of \PTM\ is a computation step denoted by $w \xrightarrow{w_{i}/w_{o}} w' $, that starts with $w$ and ends with  $w'$ on the work-tape and transforms $w_{i}$ in $w_{o}$;

\item a \PTM\ $computation$ is a sequence of {\sl macrosteps}. 
\end{itemize}

\noindent $ w \xrightarrow{w_{i}/\mu} s_{div} $ denotes a macrostep of a computation that diverges (that is a non-terminating computation);  $s_{div}$ is a particular state where each divergent computations falls, and $\mu$ is special output symbol signifying divergence; $\mu \notin \Sigma$.\\

 Moreover, the definition of the {\sl interactive transition system} (\ITS) equipped with three notions of behavioral equivalence -- \ITS\ isomorphism, interactive bisimulation, and interaction stream equivalence --  allows them to determine the \PTMs\ equivalence.
 
\begin{defn}[Interactive transition system] Given a finite alphabet 􏰅$\Sigma$ not containing $\mu$􏰈,  an \ITS\ over $\Sigma$ is a triple $\langle S, \m, r \rangle$ where 
\begin{itemize}
\item $S \subseteq \Sigma^{*} \cup \{s_{div}\} $  is the set of states; 
\item $\m\ \subseteq   S \times \Sigma^{*}􏰅 \times S \times (\Sigma^{*}􏰅 \cup \{\mu\}􏰈) )$ is the transition relation;
\item $r$ denotes the initial state.
\end{itemize} 
It is assumed that all the states in $S$ are reachable from $r$. Intuitively, a transition $\langle s, w_{i}, s', w_{o} \rangle$ of an \ITS\ states that while the machine is in the state $s$ and having received the input string $w_{i}$ from the \environment, the \ITS\ transits to state $s'$ and output $w_{o}$.
\end{defn} \label{def:ITS}

%$\forall w\in S$ and $w_{i} \in \Sigma^{*}$ whenever $w \xrightarrow{w_{i}/w_{o}} s_{div} $ then $w_{o}=\mu$ and whenever $s_{div} \xrightarrow{w_{i}/w_{o}} w$, then $w=s_{div}$ and $w_{o}=\mu$.\\

%\noindent The \ITS $\mathcal{S}= \langle S, \m, r \rangle$ associated with a \PTM  $\mathcal{M}= \langle Q, \Sigma, \C, s_{0}    \rangle$ will have
%\begin{itemize}
%\item the initial state is $r=s_{0}$.
%\item $[\dot]$: Q \rightarrow \Sigma^{*} \cup \{s_{div}\}$ the set of states  reachable from $r$ in $\mathcal{M}$ by macrosteps.
%\item the $(\Sigma^{*}􏰅 \times (\Sigma^{*}􏰅 \cup \{\mu\}􏰈) )$ transition relation is defined for all reachable $w, w' \in \Sigma^{*} \cup \{s_{div}\}$, and for all $w_{i}\in \Sigma^{*}$ and $w_{o}\in \Sigma^{*} \cup \{\mu\}$ by $w \xrightarrow{w_{i}/w_{o}} w'$  if this is a macrostep associated to $\mathcal{M}$.
%\end{itemize}

Unfortunately, the sake of space economy forced to omit most of the results;  we only recall Theorem 24, Theorem 32  and Thesis 50 (in the sequel renumbered Theorem 1, Theorem 2 and Thesis 1, respectively) and address the reader eager for more information to the original article~\cite{Goldin2004}.

\begin{thm}
The structures $\langle \mathbb{M}, =_{ms }\rangle$ and $\langle \mathbb{T}, =_{iso} \rangle$ are isomorphic.\label{th-iso}
\end{thm}

\noindent Theorem~\ref{th-iso} states that there exists a one-to-one correspondence between the class of \PTMs, denoted by $\mathbb{M}$ up to macrostep equivalence, denoted by $=_{ms}$, and the class of $ITSs$, denoted by $\mathbb{T}$ up to isomorphism, denoted by $=_{iso}$.

\begin{thm}  If a \PTM\ $\MM$ has unbounded nondeterminism, then $\MM$ diverges. \label{th-div}
\end{thm}

\noindent Theorem~\ref{th-div}  states that a \PTM\ $\MM$ diverges if there exists some $w\in \textsf{reach}({\MM})$, $w_{i}\in \Sigma^{*}$ such that  there is an infinite number of $ w_{o} \in \Sigma^{*} \cup \{\mu\}$, $w' \in \Sigma^{*} \cup \{s_{div}\}$, such that $w \xrightarrow{w_{i}/w_{o}} w' $. 

\begin{thesis} Any sequential interactive computation can be performed by a \PTM. \label{thesis}
\end{thesis}

%\medskip
Like the Church-Turing Thesis, Thesis~\ref{thesis}  cannot be proved. Informally, each step of a sequential interactive computation, corresponding to a single input/output-pair transition, is algorithmic. Therefore, by the Church-Turing Thesis, each step is computable by a \TM. A sequential interactive computation may be history-dependent, so state information must be maintained between steps. A \PTM\ is just a \TM\ that maintains state information on its work-tape between two steps. Thus, any sequential interaction machine can be simulated by a \PTM\ with possibly infinite input.

\paragraph*{The \PTM\ environment.}
In her earlier work~\cite{Goldin2000}, D. Goldin proposed a notion of \environment\ to highlight that the class of behaviors captured by the \TM, the class of algorithmic behaviors, is different from that represented by the \PTM\ model, the {\sl sequential interactive behaviors}. The conceptualization of  the \environment\ provides the observational characterization of \PTM\ behaviors given by the input-output streams. 
In fact, given two different environments $\E_{1}$ and $\E_{2}$ and a \PTM\ machine $\MM$, the behavior of $\MM$ observed by interacting with an environment $\E_{1}$ can be different if observed by interacting with $\E_{2}$. 
 Also, given two machines $\MM_{1}$ and $\MM_{2}$ and one environment $\E$, if the behaviors of the two machines are equal (one can be reduced to the other), they must be equivalent in $\E$. This claim gives the go-ahead to Theorem~\ref{th-env}. Any environment $\E$ induces a partitioning of $\M$ into equivalence classes whose members appear behaviorally equivalent in $\E$; the set of equivalence classes is denoted by $\beta_{o}$.  
Indeed, the equivalences of the behaviors of two \PTMs\ can be expressed by the language represented in the set of all interaction streams. 

\medskip
Let $\B(\MM)$ denote the operator that extracts the behavior of a given machine $\MM$, and $\E(\MM)$  a mapping that associates any machine $\MM$ to the class of the behaviors feasible for the environment $\E$. 
Therefore, each machine can be classified by analyzing its interaction streams with the two operators, $\B$ and $\E$.

\begin{defn}[Environment] Given a class $\M$ of \PTMs\ and a set of suitable domains  $\bb$, that is the set of equivalence classes of feasible behaviours. An \emph{environment} $\E$ is a mapping from machines to some domains $\E : \M \rightarrow \bb $
and the following property holds: 

\[ \forall \, \MM_{1}, \MM_{2} \in \M, \,\,\,  \text{if} \, \,\, \B(\MM_{1}) = \B(\MM_{2}) \, \,\, \text{then} \, \,\, \E(\MM_{1}) = \E(\MM_{2}) \]

 \medskip \noindent When $\E(\MM_{1}) \neq \E(\MM_{2})$, we say that $\MM_{1}$ and $\MM_{2}$ are \emph {distinguishable} in $\E$; otherwise, we say that $\MM_{1}$ and $\MM_{2}$ appear \emph{equivalent} in $\E$.
 \end{defn}
 
\begin{thm}
Let $\Theta$ denote the set of all possible environments. The environments in $\Theta$ induce an infinite expressiveness hierarchy of \PTM\ behaviors, with \TM\ behaviors at the bottom of the hierarchy. \label{th-env}
\end{thm}

So far, we have assumed that all the input streams are all feasible. However, this is not a reasonable assumption for the context in which interactive machines normally run. Typically an environment can be constrained and limited by some obstructions when generating the output streams. In our view, this is the case where the space of all possible configurations lies on a topological space with not trivial topology. In order to contribute to this theory, in the following we will tackle the issue of specifying these constraints, and relating them to the \PTM\ model. 

%
%%%%%%%%%%%%%%%%%%%%%
\section{Topological Interpretation of Interactive Computation}
\label{sec-topo}
\tdplotsetmaincoords{60}{0}
\tikzset{declare function={torusx(\u,\v,\R,\r)=cos(\u)*(\R + \r*cos(\v)); 
torusy(\u,\v,\R,\r)=(\R + \r*cos(\v))*sin(\u);
torusz(\u,\v,\R,\r)=\r*sin(\v);
vcrit1(\u,\th)=atan(tan(\th)*sin(\u));% first critical v value
vcrit2(\u,\th)=180+atan(tan(\th)*sin(\u));% second critical v value
disc(\th,\R,\r)=((pow(\r,2)-pow(\R,2))*pow(cot(\th),2)+% 
pow(\r,2)*(2+pow(tan(\th),2)))/pow(\R,2);% discriminant
umax(\th,\R,\r)=ifthenelse(disc(\th,\R,\r)>0,asin(sqrt(abs(disc(\th,\R,\r)))),0);
}}

\paragraph*{Topological environment}

This section deals with the notion of {\sl topological environment} as an integral part of the model of topological computation. In a classical \TM\ the environment is not represented (Def:\ref{def:TM}), whereas in a \PTM\ the environment is a mapping between the class of \PTMs\ and their feasible domains. As described above the two functions $\B$ and $\E$ permit to identified the behavior of a  \PTM\ machine by observing its stream of interactions. In this case the environment $\E$ is a static mapping that associates machines with an equivalent behavior $\B(\MM)$ to the same equivalence class. In this case the environment plays the role of an observer. In our approach the environment is part of the system that evolves together with the behavior of the machine over time step $i$. The environment constrains the behavior of a machine \PTM\ so as the output generated by the machine affects the evolution of the environment. 

To detect dynamic changes in the environment, we propose to define a dynamic analysis of the set of all the interactions streams available at any single \PTM\ computation step $i$. Since interaction streams are infinite sequences of pairs of the form $(w_{i}, w_{o})$ representing the input and output strings of \PTMs\ computation step $i$, we use the set $\C$  of \PTM\ configurations to represent them.

The resulting model of computation consists of two components entangled and coexisting during the interactive computation, a functional unit of computation and a {\sl self-organizing memory}. 

In our model, the infinite input of the \PTM\ should be seen as a feedback loop of a dynamic system. Its functional behavior is represented by a class $\T$ of \ITS\ constrained by the information contained in the self-organizing memory associated with the  notion of topological environment. The data structure used to store information is the simplicial complex $\SC$, that is a topological space $\TS$ constructed over the set of \PTM\ configuration $\C$. The $\SC$ is equipped with a finite presentation in terms of homology groups whose relations are fully representable. In this view the \PTM\ functional behavior can be determined by $\SC$ modulo \ITS\ isomorphism. We operate in a discrete setting where full information about topological space is inherent in their simplicial representation. Appendix~1 provides some useful definitions for algebraic and computational topology. 
 
\begin{defn}[Topological environment] 
Given the set of \PTM\ configurations $\C_{i}$ available at a given time $i$, the \emph{topological environment} is the simplicial complex $\TS_{\C_{i}}$ constructed over $\C_{i}$.
\end{defn}

The topological environment $\TS_{\C}$, as any topological space is equipped with a set of invariants that are important to understand the characteristics of the space.  For the sake of simplicity we will refer to topological space as a continuous space. The $n$-dimensional holes, the language of paths, the homology and the genus are topological invariants. The $n$-dimensional holes are determined during the process of filtration, called persistent homology, that is used to construct a topological space starting from a set of points. The numbers of holes and their associated dimensions are determined by the {\em homology} structure fully represented by the homology groups associated with a {\em topological space}.
Also the homology is a topological invariant of the space, it is always preserved by homeomorphisms of the space. 

\input{path-cyclic}

A {\sl path} in a topological space $\TS$ is a continuous function  $f: [0,1] \rightarrow \TS$ from the unit interval to $\TS$. Paths are oriented, thus $f(0)$ is the starting point and $f(1)$ is the end-point, if we label the starting point $v$ and the end-point $v'$, we call $f$ a path from $v$ to $v'$ as shown in Figure~\ref{fig:path}-(a). Two paths $a$ and $b$, that is two continuous functions, from a topological space $\TS$ to a topological space $\TS'$ are homotopic if one can be continuously deformed into the other. Being homotopic is an equivalence relation on the set of all continuous functions from $\TS$ to $\TS'$. The homotopy relation is compatible with function composition. 
\input{path-gen}

Therefore, it is interesting to study the effect of the existence of holes (at any dimension) in a topological space $\TS$ (for simplicity the discussion is made thinking of $\TS$ as a 2D surface) built from the space of configurations $\C$ where a sequential interactive computation takes place as a {\em sequential composition of paths}. Figures~\ref{fig:path}-(b) and -(c) show the composition of two paths $a$ and $b$, and the  proof that they are not homotopic, respectively.  Given two-cycle paths, $a$ and $b$, with a point in common in $x$,  if the composition of the two paths $ab$ or $ba$ is not commutative, the two composed paths are not equivalent. In this case, the two cycle paths, $a$ and $b$ can be considered the generators of a topological space with one 2-dimensional hole, as shown in Figure~\ref{fig:homotopy}. Each generator represents a distinct class of paths, $[a]$ those going around the neck, and $[b]$ those around the belt of the torus, respectively. 
%We denote the class of algorithms $[A]$ associated to the same function $f$, and with $f_{[A]}$ the general function defined over any deformation of the space $\TS$.

\begin{figure}[h]
\centering
% keynote pictures
\includegraphics [width=4cm]{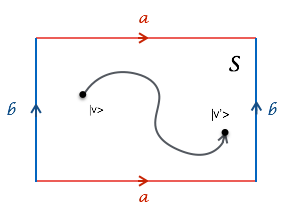} \hspace{0.5cm}
\includegraphics [width=4cm]{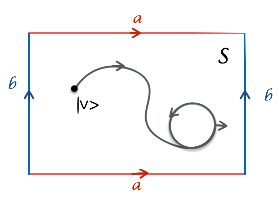} \hspace{0.5cm}
\includegraphics [width=4cm]{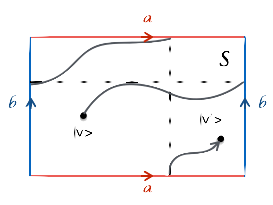} 
\caption{a) successful computation, b) computation with an infinite loop, c) computation with a deadlock.} \label{fig:vs-oriented}
\end{figure}

\paragraph*{Computable functions and topological space.} We start taking into account those classes of problems whose computable functions are defined over a space $\TS$ endowed with a trivial topology, and it is a Vector Space.
Figure~\ref{fig:vs-oriented} shows how an algorithmic computation $A$ associated with the function $f_{A}: \TS \rightarrow \TS$, evolves over  $\TS,$ representing the space of the states. Each state $v$ is defined by a vector that moves over  $\TS$ driven by the configurations of the \TM.  
\begin{figure}[t]
\centering
%keynote picture
\includegraphics [width=4cm]{deadlock} \hspace{0.3cm}
\includegraphics [width=2.5cm]{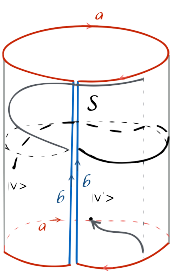} \hspace{0.3cm}
\input{torus-path}
\caption{A deadlocked computation on the plane may successes over a space with non-trivial topology.} \label{fig:torus}
\end{figure}
In Figure~\ref{fig:vs-oriented}, from left to right, the first two pictures represent a successful computation and a computation with an infinite loop, respectively. When the algorithm moves the vector towards a {\em boundary}, see the last picture, the computation is deadlocked. This happens because $\TS$ has not been defined globally. In fact, the boundary breaks the translational symmetry. If we allow the boundary to disappear by adding an extra-relation, global in nature, we obtain a global topology that is not trivial -- the space is characterized by a not empty set of n-dimensional holes ($n\geq2$). Figure~\ref{fig:torus} shows how the computation with a deadlock on the plane could have succeeded if the manifold of the space is a torus. 

\begin{figure}[h]
\centering
%Tzk picture
%%%%%% from plane to torus
\input{plane-dash}
A)
\includegraphics [width=4cm]{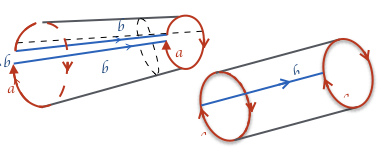}
B)
\includegraphics [width=4cm]{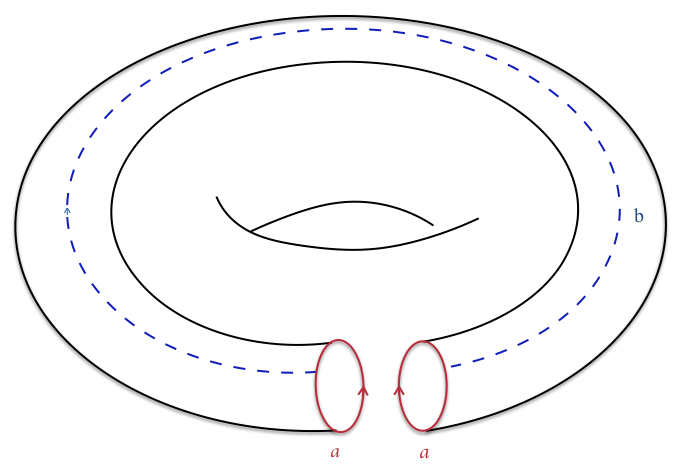}
C)
%
%%%% torus and 2 generators
\input{torus}
D)
\caption{The pictures (A--D) summarize the main steps to transform a space  $\TS$ of \PTM\ into a topological space $ \SC$. The construction  is obtained by gluing together -- put in relation -- the two boundaries of the space  $\TS$,  $a$ and $b$ respectively, which become the generators $\mathbf{a}$ and $\mathbf{b}$ of the new space $\SC$.  The topological space $\SC$, finite but not limited, naturally supports the notion of the environment of \PTM.} \label{fig:topo-space}
\end{figure}

\input{alpha-gamma}

Figure~\ref{fig:topo-space} shows how we can transform a rectangle, 2-dimensional space $\mathcal{S}$ homomorphic to 2-manifold with boundary,  into a cylinder and then into a torus by adding two relations among the generators of the manifold $\mathcal{P}$ that will be proved to be {\em without boundary}. 

Hence, we proceed to analyze those classes of problems whose computable functions are defined over a space $\mathcal{S}$ endowed with non-trivial topology. The class of functions $F_{\mathcal{S}}$ effectively computable over a space $\mathcal{S}$,  and for each single function $f_{A}\in F_{\mathcal{S}}$ and a couple of points $v,v' \in \mathcal{S}$, we associate a computation $f_{A}(v)=v'$, as a path that connects the two points $v$ and $v'$ in the space $\mathcal{S}$. The path can be semantically interpreted as an interaction stream. 

In Figure~\ref{fig:paths}, the first two pictures from left to right, show that a close path $\pi$ in a surface that starts and ends to a fixed point $P_{B}$ is homotopic to 0; it means that any $\pi$ can be reduced to the point $P_{B}$. The class of behavioral equivalence to $\tau$ denoted by $[\pi]$ belongs to space or subspace space with trivial topology $g=0$ ($g$ is the {\sl genus}). The other pictures show irreducible paths belonging to space with a topological genus $g\neq 0$. E.g. if $g=1$, i.e. is a torus there are three different classes of behaviors: i) the set of closed paths homotopic to $0$. In this case, we are given a local interpretation and we are not aware that at the global level the {\sl genus} can be different from $0$; ii) the set of closed paths homotopic to the first generator $a$ of the homology group of the topological space $\TS$. The cycle fixed on the base point $P_{B}$ can be used to reduce any path going around the belt of the torus to $a$ by a continuous deformation; iii) similar to the previous set, but the paths are homotopic to the second generator $b$ of $\TS$. The cycle fixed on the base point $P_{B}$ goes around the belt of the torus. The last picture shows the composition of paths.

The interpretation of interaction streams over a $\TS_{\C}$ is indeed nothing but its identification with an element of the path algebra corresponding to a quiver representation of the transformation group $\mathcal{G}$ of $\TS$, say $\Q$  (or, more generally, a set of quivers, over some arbitrary ring). The different ways to reach any point $p\in\C$ from $P_{B}$ generate a path algebra $\A$ whose elements are describable words in a language $\La$. Any point of $\C$ can be related to any other point by a group element. By selecting a point $p_{0}$ of $\C$ as a unique base point, every point of $\C$ is in one-to-one correspondence with an element of such group $\mathcal{G_{MC}} \approx \mathfrak{MCG}$, the simplicial analog of the mapping class group. $\mathcal{G_{MC}}$ is a group of transformations which do not change the information hidden in the data, such as the group of diffeomorfisms that do not change the topology of the base space. $\mathfrak{MCG}$ is an algebraic invariant of a topological space, that is a discrete group of symmetries. Since the algebras manipulate the data, the transformations applied to space are $'$processes$'$ carried on through the fiber, which is the representation space of the process algebra. Whenever $\Q$ can give the representation of the algebra, the algebra  can be exponentiated to a group $\mathcal{G_{AP}}$ and t a gauge group. We have now all the ingredients for defining a fiber bundle enriched with a group $\mathcal{G}=\mathcal{G_{AP}} \wedge \mathcal{G_{MC}}$, called {\sl gauge group}, (see Figure~\ref{img:topo}). Summarizing, fiber bundle is the mathematical structure that allows us to represent computation and its context (the environment) as a unique model. In terms of \TM, the context represents the {\sl transition function}, also called the {\sl functional matrix}.

%superstatement
\begin{center}{\sl While the algorithmic aspect of a computation expresses the effectiveness of the computation, the topology provides a global characterization of the environment.} \end{center}

\noindent Both the computation and the environment can be represented as groups (algebras), and their {\sl interaction} is captured as the set of accessible transformations of  the semi-direct product of the two groups, carrying constrained by the restrictions imposed by topology. Incidentally, it is this set of constraints together with the semidirect product structure that implies the non-linearity of the process. 

\begin{defn}[Topological Turing machine] A Topological Turing machine (\ToM) is a group $\G$ consisting of all interaction streams generated by the group of \PTMs\ entangled with the group of all transformations of the topological space $\TS_{\C}$ preserving the topology. Formally $\G=\G_{AP} \wedge \G_{MC}$, where $\G_{AP}$ is the group of \PTMs\ and $\G_{MC}$ the simplicial analog of the mapping class group.

\end{defn}

\begin{prp} If $\G$ is automatic, the associated language $\La$ is regular. Since the representations of $\G$ can then be constructed in terms of quivers $\Q$ with relations induced by the corresponding path algebra induced by \PTMs, the syntax of $\La$ is fully contained in $\T$ and its semantics in $\M$.
\end{prp}

\begin{defn}[Constrained interactive computation] An interactive computation is constrained if it is defined over a topological space $\SC$ and it is an element of the language of paths of $\SC$.
\end{defn}

\begin{thm} Any \emph {constrained interactive computation} is an effective computation for a \ToM.

\end{thm}

\begin{thesis} Any concurrent computation can be performed by a \ToM.

\end{thesis}
\section{Final remarks}
\label{sec:conclusions}

In 2013, Terry Tao in his blog~\cite{TerryTao2012} posted this question: if there is any computable group $G$ which is $''$Turing complete$''$ in the sense that the halting problem for any Turing machine can be converted into a question of the above form. In other words, there would be {\bf an algorithm} which, when given a Turing machine $T$, would return (in a finite time) a pair $x_{T}, y_{T}$  of elements of  $G$  with the property that  $x_{T}, y_{T}$ generate a free group in  $G$ if and only if  $T$ does not halt in finite time. Or more informally: {\sl can a $'$group$'$ be a universal Turing machine?}

\section*{Acknowledgements}
E. M. thanks Mario Rasetti for bringing her to conceive a new way of thinking about computer science and for numerous and lively discussions on topics related to this article; and Samson Abramsky with his group for insightful conversations on the topological interpretation of contextuality and contextual semantics.  E.~M. and A.W. thank the anonymous referees for suggesting many significant improvements.\\
{\em Funding statements.} We acknowledge the financial support of the Future and Emerging Technologies (FET) programme within the Seventh Framework Programme (FP7) for Research of the European Commission, under the FP7 FET-Proactive Call 8 - DyMCS, Grant Agreement TOPDRIM, number FP7-ICT-318121.

%\bibliographystyle{abbrv}
%\bibliography{biblio}

\newpage
\section*{Appendix 1: Definitions of Algebraic and Computational Topology}
%
%\input{src/appendix}
%\label{sec:appendix}
{\small
\begin{defn}Topology\\
A \textit{topology} on a set X is a family $T \subseteq 2^X $ such that
\begin{itemize}
\item[-]  If $S_{1}, S_{2} \in T,$ then $S_{1} \cap S_{2} \in T$ (equivalent to: If $S_{1}, S_{2}, \dots, S_{n}  \in T$ then  $\cap_{i=1}^{n} S_{i} \in T$) 
\item[-]  If $\{ S_{j} | j \in J  \} \subseteq T, $ then $\cup_{j \in J}S_{j} \in T.$
\item[-]  $\emptyset,  X \in T.$ 
\end{itemize}
\end{defn}

\begin{defn}Topological spaces\\
The pair (X,T) of a set X and a topology T is a \textit{topological space}. We will often use the notation $\mathbb X$ for a topological space X, with T being understood.
\end{defn}

\begin{defn}Simplices\\
Let \textit{$u_{0},u_{1}, ..., u_{k}$} be points in $\mathbb{R}^{d}.$ A point x =$\sum_{i=0}^{k}\lambda_{i}u_{i}$ is an \textit{affine combination} of the $u_{i}$ if the $\lambda_{i}$ sum to 1. The \textit{affine hull} is the set of affine combinations. It is a \textit{k-plane} if the k+1 points are affinely independent by which we mean that any two affine combinations, x=$\sum_{i=0}^{k}\lambda_{i}u_{i}$ and y =$\sum_{i=0}^{k}\mu_{i}u_{i}$ are the same iff $\lambda_{i}=\mu_{i}$ for all \textit{i}. The k+1 points are affinely independent iff the k vectors $u_{i} \dots u_{0}$, for $1\leq i \leq k,$ are linearly independent. In $\mathbb{R}^{d}$ we can have at most \textit{d} linearly independent vectors and therefore at most \textit{d+1} affinely independent points.\\
 \textit{k-simplex} is the convex hull of k+1 affinely independent points, $\sigma=\{u_{0},u_{1},u_{2},...u_{k}\}$. Its \textit{dimension} is $dim \sigma = k$. Any subset of affinely independent points is again independent and therefore also defines a simplex of lower dimension. \end{defn}

\begin{defn}Face\\
A \textit{face} of $\sigma$ is the convex hull of a non-empty subset of the $u_{i}$ and it is \textit{proper} if the subset is not the entire set. We sometimes write $\tau \leq \sigma $ if $\tau$ is a face and $\tau < \sigma$ if it is a proper face of $\sigma.$ Since a set of k+1 has $2^{k+1}$ subsets, including empty set, $\sigma $ has $2^{k+1} -1$ faces, all of which are proper except for $\sigma$ itself. The \textit{boundary} of $\sigma$, denoted as \textit{$\rm bd \sigma$}, is the union of all proper faces, and the \textit{interior} is everything else.
\end{defn}

\begin{defn}Simplicial complexes\\
A \textit{simplical complex} is a finite collection of simplices K such that $\sigma \in$ K and $\tau \in$ K, and $\sigma, \sigma_{0} \in$ K implies $\sigma \cap \sigma_{0}$ is either empty or a face of both.
\end{defn}

\begin{defn}Filtration\\
A \textit{filtration} of a complex K is a nested sequence of subcomplex, $\emptyset=K^{0} \subseteq K^{1}\subseteq K^{2}\subseteq....\subseteq K^{m} = K$. We call a complex K with a filtration a filtered complex.
\end{defn}

\begin{defn}Chain group\\
The \textit{k-th chain group} of a simplicial complex K is $\langle C_{k}(K),+\rangle$, let $\mathbb{F}$ be a field. The $\mathbb{F}-$linear space on the oriented \textit{k-simplices}, where $[\sigma]=-[\tau]$ if $\sigma=\tau$ and $\sigma$ and $\tau$ have different orientations. An element of $C_{k}(K)$ is a \textit{k-chain}, $\sum_{q}n_{q}[\sigma_{q}], n_{q} \in \mathbb Z,\sigma_{q} \in K$.
\end{defn}

\begin{figure}[h]
\centering
\includegraphics[width=7cm]{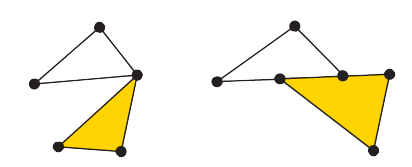}
\centerline{A simplicial complex (left) and not valid simplicial complex (right).}
\includegraphics[width=6cm]{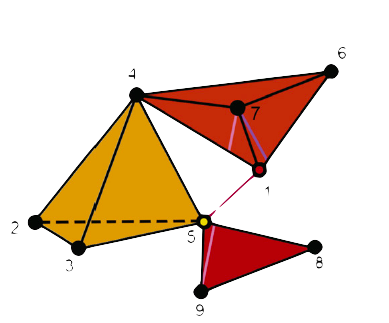}
\includegraphics[width=6cm]{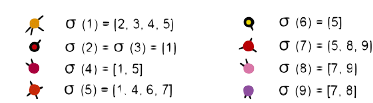}
\centerline{A simplicial complex and its simplices.}
\end{figure}
\begin{defn}Boundary homomorphism\\
Let K be a simplicial complex and $\sigma \in K, \sigma=[v_{0},v_{1},...,v_{k}]$ The boundary homomorphism $\partial_{k}:C_{k}(K)\rightarrow C_{k-1}(K)$ is \,\,\,
$\partial_{k}\sigma = \sum_{i}(-1)^{i}[v_{0},v_{1},...,\widehat{v_{i}},... ,v_{n}]$ where $\widehat{v_{i}}$ indicates that $v_{i}$ is deleted from the sequence.
\end{defn}
\begin{defn}Cycle and boundary\\
The \textit{k-th cycle group} is $Z_{k}=ker\partial_{k}$. A chain that is an element of $Z_{k}$ is a k-cycle. The k-th boundary group is $B_{k}=im\partial_{k+1}.$ A chain that is an element of $B_{k}$ is a k-boundary. We also call boundaries \textit{bounding cycles} and cycles not in $B_{k}$ \textit{nonbounding} cycles.
\end{defn}

\begin{defn}Homology group\\
The \textit{k-th homology group} is \,\, 
$H_{k} = Z_{k}/B_{k} = ker\partial_{k}/im\partial_{k+1}.$\\
If $z_{1}=z_{2}+B_{k},z_{1},z_{2} \in Z_{k},$ we say $z_{1}$ and $z_{2}$ are homologous and denote it with $z_{1} \sim z_{2}$
\end{defn}

\begin{defn}k-th Betti number\\
The \textit{k-th Betti number} $\rm B_{k}$ of a simplicial complex K is the dimension of the \textit{k-th} homology group of K. Informally, $\beta_0$ is the number of connected components, $\beta_1$ is the number of two-dimensional holes or "handles" and $\beta_2$ is the number of three-dimensional holes or "voids" etc\dots.
\end{defn}

\begin{defn}Invariant\\
A topological \textit{invariant} is a property of a topological space which is invariant under homeomorphisms. Betti numbers are topological invariants.
\end{defn}

\begin{defn}Genus\\
The genus is a topological invariant of a close (oriented) surface. The connected sum of $g$ tori is called a surface with \textit{genus} $g$. \textit{genus} refers to how many $'$holes$'$ the donut surface has.\\ 
As an example, a torus is homeomorphic to a sphere with a handle. Both of them have just one hole (handle). The sphere has $g=0$ and the torus has $g=1$. 
\end{defn}
}

\end{document}

%% file: path-cyclic.tex
\begin{figure}[h]
\centering
% tikz pictures
%%% cyclic path
\begin{tikzpicture}[scale=0.55,tdplot_main_coords]
 \draw[color=red,decoration={markings, mark=at position 0.6 with {\arrow[very thick]{<}}},
    postaction={decorate}] (-15cm,0cm) circle(2cm);
 \draw[color=blue,decoration={markings, mark=at position -0.1 with {\arrow[very thick]{>}}},
    postaction={decorate}] (-11cm,0cm) circle(2cm);
 \draw [color=red](-17,2.8) node {$a$};
\draw [color=blue] (-8.8,2.0) node {$b$};
\draw (-13,-2.8) node {$x$};
\draw [thick](-13,-0) node {$\bullet$};
\end{tikzpicture}
%%%%% -- close path
\begin{tikzpicture}[scale=0.3,tdplot_main_coords]
\draw [color=red,thick](-17, -10) arc (30:150:7cm);
\draw [color=red](-22.5,-3) node {$>$};
\draw [color=blue,thick](-17, -10) arc (-30:-150:7cm);
\draw [color=blue](-23,-17) node {$>$};
 \draw (-30,-10) node {$|v\rangle$};
\draw (-16,-10) node {$|v'\rangle$};
\draw [thick](-29,-10) node {$\bullet$};
\draw [thick](-17,-10) node {$\bullet$};

\draw [color=red](-24,-2) node {$a$};
\draw [color=blue] (-22,-18) node {$b$};
\end{tikzpicture}\vspace{0.5cm}
%

%%%% nohomotpic path
\begin{tikzpicture}[scale=0.3,tdplot_main_coords]
% red
\draw [thick](-29.1,-10.2) node {$\bullet$};
\draw [thick](-17,-10.3) node {$\bullet$};
\draw [thick](-4.9,-10.3) node {$\bullet$};
% blu
\draw [thick](-29.4,-14.3) node {$\bullet$};
\draw [thick](-17,-19.8) node {$\bullet$};
\draw [thick](-4.5,-14.3) node {$\bullet$};
\draw [color=red,thick, <-](-17, -10) arc (30:150:7cm);
%\draw [thick, color=red](-21,-3.6) node {$>$};
\draw [color=blue,thick, ->](-17, -10) arc (150:30:7cm);
\draw [color=red,thick, ->](-17, -20) arc (-140:-15:7.2cm);
\draw [color=blue,thick, <-](-17, -20) arc (-45:-160:7.5cm);
%
% red
\draw [thick](-24,-3.1) node {$\bullet$};
\draw [thick](-24,-24) node {$\bullet$};
%
% blu
\draw [thick](-10,-3.1) node {$\bullet$};
\draw [thick](-10,-24.9) node {$\bullet$};
\draw [thick, dashed](-24,-3.1) -- (-10,-24.9);
\draw [thick, dashed](-24,-24) --(-10,-3.1);
\draw (-29.6,-10) node {$x$};
\draw (-17,-8.5) node {$x$};
\draw (-4.5,-10) node {$x$};
\draw (-29.9,-14.3) node {$x$};
\draw (-17,-21.5) node {$x$};
\draw (-4,-14.7) node {$x$};
 \draw (-24,-1.5) node {$|v\rangle$};
\draw (-10,-1.5) node {$|v'\rangle$};
 \draw (-24,-25.6) node {$|v'\rangle$};
\draw (-10,-26.5) node {$|v\rangle$};
\draw [color=red](-26,-3) node {$a$};
\draw [color=blue] (-8.5,-2.5) node {$b$};
\draw [color=blue](-26,-24) node {$b$};
\draw [color=red] (-8,-25) node {$a$};

\end{tikzpicture}

%%TzK picture
%\includegraphics [width=4cm]{img/path-cyclic} \hspace{0.5cm}
%\includegraphics [width=4cm]{img/path-closed} \hspace{0.5cm}
%\includegraphics [width=5cm]{img/path-nohomo-proof} 
\caption{a) homotopic paths $a\sim b$; b) composition of paths $ab$; c) not homotopic paths $ab \nsim ba$ } 
\label{fig:path}
\end{figure}

%% file: path-gen.tex
\begin{figure}[h]
\centering
%Tizk picture 

\begin{tikzpicture}[scale=0.5,tdplot_main_coords]

 \draw[thick,color=red,decoration={markings, mark=at position 0.6 with {\arrow[very thick]{<}}},
    postaction={decorate}] (-15cm,0cm) circle(2cm);
 \draw[thick,color=blue,decoration={markings, mark=at position -0.1 with {\arrow[very thick]{>}}},
    postaction={decorate}] (-11cm,0cm) circle(2cm);
\draw [color=red](-17,2.8) node {$a$};
\draw [color=blue](-8.8,2.0) node {$b$};
\draw (-13,-2.8) node {$x$};
\draw (-13,-0.5) node {$\bullet$};
%\end{tikzpicture}
%
% torus 3D with two generators
%\begin{tikzpicture}[scale=1,tdplot_main_coords]
\pgfmathsetmacro{\R}{4}
\pgfmathsetmacro{\r}{1.5}
 \draw[thick,even odd rule,fill opacity=0.2] plot[variable=\x,domain=0:360,smooth,samples=71]
 ({torusx(\x,vcrit1(\x,\tdplotmaintheta),\R,\r)},
 {torusy(\x,vcrit1(\x,\tdplotmaintheta),\R,\r)},
 {torusz(\x,vcrit1(\x,\tdplotmaintheta),\R,\r)}) 
 plot[variable=\x,
 domain={-180+umax(\tdplotmaintheta,\R,\r)}:{-umax(\tdplotmaintheta,\R,\r)},smooth,samples=51]
 ({torusx(\x,vcrit2(\x,\tdplotmaintheta),\R,\r)},
 {torusy(\x,vcrit2(\x,\tdplotmaintheta),\R,\r)},
 {torusz(\x,vcrit2(\x,\tdplotmaintheta),\R,\r)})
 plot[variable=\x,
 domain={umax(\tdplotmaintheta,\R,\r)}:{180-umax(\tdplotmaintheta,\R,\r)},smooth,samples=51]
 ({torusx(\x,vcrit2(\x,\tdplotmaintheta),\R,\r)},
 {torusy(\x,vcrit2(\x,\tdplotmaintheta),\R,\r)},
 {torusz(\x,vcrit2(\x,\tdplotmaintheta),\R,\r)});
 % throat
 \draw[thick] plot[variable=\x,
 domain={-180+umax(\tdplotmaintheta,\R,\r)/2}:{-umax(\tdplotmaintheta,\R,\r)/2},smooth,samples=51]
 ({torusx(\x,vcrit2(\x,\tdplotmaintheta),\R,\r)},
 {torusy(\x,vcrit2(\x,\tdplotmaintheta),\R,\r)},
 {torusz(\x,vcrit2(\x,\tdplotmaintheta),\R,\r)});
 \foreach \X  in {300}  
 {\draw[color=red,thick,dashed] 
  plot[smooth,variable=\x,domain={360+vcrit1(\X,\tdplotmaintheta)}:{vcrit2(\X,\tdplotmaintheta)},samples=71]   
 ({torusx(\X,\x,\R,\r)},{torusy(\X,\x,\R,\r)},{torusz(\X,\x,\R,\r)});
 
 \draw[color=red,thick] 
  plot[smooth,variable=\x,domain={vcrit2(\X,\tdplotmaintheta)}:{vcrit1(\X,\tdplotmaintheta)},samples=71]   
 ({torusx(\X,\x,\R,\r)},{torusy(\X,\x,\R,\r)},{torusz(\X,\x,\R,\r)});
 \draw[color=red,thick,-latex] 
  plot[smooth,variable=\x,domain={vcrit1(\X,\tdplotmaintheta)}:40,samples=71]   
 ({torusx(\X,\x,\R,\r)},{torusy(\X,\x,\R,\r)},{torusz(\X,\x,\R,\r)});
 }
 % neck
 \draw[color=blue,thick,-latex] plot[smooth,variable=\x,domain=00:360,samples=71]   
 ({torusx(\x,90,\R,\r)},
 {torusy(\x,90,\R,\r)},
 {torusz(\x,90,\R,\r)}); 
 \draw [color=blue,thick](4.3,1.0) node {$b$};
\draw [color=red,thick](1.0,-3.0) node {$a$};
\draw [thick](2.6,-1.3) node {$x$};
\draw [thick](2.1,-1.0) node {$\bullet$};
\end{tikzpicture}

%\includegraphics [width=12cm]{img/path+gen} \\
%Keynote pictures
%\includegraphics [width=4cm]{img/path}
%\includegraphics [width=6cm]{img/torusX-B}
\caption{ From cycling paths to generators of a space $\TS$} \label{fig:homotopy}
\end{figure}

%% file: torus-path.tex
\begin{tikzpicture}[scale=0.39,tdplot_main_coords]
%------------------------ 3D torus with a complex path
\draw [thick, red](-2.1, -1.25) arc (180:90:0.4cm);
\draw [thick, red, dashed,rotate=-30](1.5,-5) arc (190:300:-2cm);
\draw [thick, red, rotate=280](.6, 2.5) arc (300:180:-1.9cm);
\draw [thick, red, dashed, rotate=-5](1.5, -.5) arc (300:180:-1.9cm);

\draw [thick, red, rotate=-50,<-](2.2, 6.5) arc (300:180:-1.5cm);
\draw[thick,fill] (3.7,1.1) circle (0.06cm);
\draw[thick,fill] (-2.1,-1.2) circle (0.06cm);

\draw[thick, blue, <-] (3.7,1.5) arc (-10:240:3.9cm and 2 cm);
\draw [thick,red](0.2, -7) arc (250:170:0.4cm);
\draw [thick,red](4.33, -4.5) arc (250:170:0.4cm);
\draw [thick,red](1.42, -1.1) arc (180:60:0.3cm);
 \draw (-2.8,-1.8) node {$|v\rangle$};
\draw (4.5,1.8) node {$|v'\rangle$};

%\begin{tikzpicture}[scale=1,tdplot_main_coords]
\pgfmathsetmacro{\R}{4}
\pgfmathsetmacro{\r}{1.5}
 \draw[thick,,even odd rule,fill opacity=0.2] plot[variable=\x,domain=0:360,smooth,samples=71]
 ({torusx(\x,vcrit1(\x,\tdplotmaintheta),\R,\r)},
 {torusy(\x,vcrit1(\x,\tdplotmaintheta),\R,\r)},
 {torusz(\x,vcrit1(\x,\tdplotmaintheta),\R,\r)}) 
 plot[variable=\x,
 domain={-180+umax(\tdplotmaintheta,\R,\r)}:{-umax(\tdplotmaintheta,\R,\r)},smooth,samples=51]
 ({torusx(\x,vcrit2(\x,\tdplotmaintheta),\R,\r)},
 {torusy(\x,vcrit2(\x,\tdplotmaintheta),\R,\r)},
 {torusz(\x,vcrit2(\x,\tdplotmaintheta),\R,\r)})
 plot[variable=\x,
 domain={umax(\tdplotmaintheta,\R,\r)}:{180-umax(\tdplotmaintheta,\R,\r)},smooth,samples=51]
 ({torusx(\x,vcrit2(\x,\tdplotmaintheta),\R,\r)},
 {torusy(\x,vcrit2(\x,\tdplotmaintheta),\R,\r)},
 {torusz(\x,vcrit2(\x,\tdplotmaintheta),\R,\r)});
 
 % holes
 \draw[thick] plot[variable=\x,
 domain={-180+umax(\tdplotmaintheta,\R,\r)/2}:{-umax(\tdplotmaintheta,\R,\r)/2},smooth,samples=51]
 ({torusx(\x,vcrit2(\x,\tdplotmaintheta),\R,\r)},
 {torusy(\x,vcrit2(\x,\tdplotmaintheta),\R,\r)},
 {torusz(\x,vcrit2(\x,\tdplotmaintheta),\R,\r)});
 \end{tikzpicture}

%% file: plane-dash.tex
\begin{tikzpicture}[scale=0.5]
%grid
%\draw [step=0.5cm] (0,0) grid (5,10);
%
\draw[color=red, thick, ->] (1,8) -- (4,8)  node[above]{$a$};
\draw[color=red,thick] (4,8) -- (7,8);
\draw[color=blue, thick] (1,8) -- (1,6) node[left]{$b$};
\draw[color=blue, thick, <-] (1,6) -- (1,4);
\draw[color=red, thick, ->] (1,4) -- (4,4) node[below]{$a$};
\draw[color=red, thick] (4,4) -- (7,4);
\draw[color=blue, thick, ->] (7,4) -- (7,6) node[right]{$b$};
\draw[color=blue, thick ] (7,6) -- (7,8);
%
% dashed lines
%
\draw[thick,dashed] (1,7) -- (7,7);
\draw[thick,dashed ] (5,4) -- (5,8);
\end{tikzpicture}

%% file: torus.tex
\begin{tikzpicture}[scale=0.4,tdplot_main_coords]
% torus 3D with two generators
%\begin{tikzpicture}[scale=1,tdplot_main_coords]
\pgfmathsetmacro{\R}{4}
\pgfmathsetmacro{\r}{1.5}
 \draw[thick,even odd rule,fill opacity=0.2] plot[variable=\x,domain=0:360,smooth,samples=71]
 ({torusx(\x,vcrit1(\x,\tdplotmaintheta),\R,\r)},
 {torusy(\x,vcrit1(\x,\tdplotmaintheta),\R,\r)},
 {torusz(\x,vcrit1(\x,\tdplotmaintheta),\R,\r)}) 
 plot[variable=\x,
 domain={-180+umax(\tdplotmaintheta,\R,\r)}:{-umax(\tdplotmaintheta,\R,\r)},smooth,samples=51]
 ({torusx(\x,vcrit2(\x,\tdplotmaintheta),\R,\r)},
 {torusy(\x,vcrit2(\x,\tdplotmaintheta),\R,\r)},
 {torusz(\x,vcrit2(\x,\tdplotmaintheta),\R,\r)})
 plot[variable=\x,
 domain={umax(\tdplotmaintheta,\R,\r)}:{180-umax(\tdplotmaintheta,\R,\r)},smooth,samples=51]
 ({torusx(\x,vcrit2(\x,\tdplotmaintheta),\R,\r)},
 {torusy(\x,vcrit2(\x,\tdplotmaintheta),\R,\r)},
 {torusz(\x,vcrit2(\x,\tdplotmaintheta),\R,\r)});
 % throat
 \draw[thick] plot[variable=\x,
 domain={-180+umax(\tdplotmaintheta,\R,\r)/2}:{-umax(\tdplotmaintheta,\R,\r)/2},smooth,samples=51]
 ({torusx(\x,vcrit2(\x,\tdplotmaintheta),\R,\r)},
 {torusy(\x,vcrit2(\x,\tdplotmaintheta),\R,\r)},
 {torusz(\x,vcrit2(\x,\tdplotmaintheta),\R,\r)});
 \foreach \X  in {300}  
 {\draw[color=red,thick,dashed] 
  plot[smooth,variable=\x,domain={360+vcrit1(\X,\tdplotmaintheta)}:{vcrit2(\X,\tdplotmaintheta)},samples=71]   
 ({torusx(\X,\x,\R,\r)},{torusy(\X,\x,\R,\r)},{torusz(\X,\x,\R,\r)});
 
 \draw[color=red,thick] 
  plot[smooth,variable=\x,domain={vcrit2(\X,\tdplotmaintheta)}:{vcrit1(\X,\tdplotmaintheta)},samples=71]   
 ({torusx(\X,\x,\R,\r)},{torusy(\X,\x,\R,\r)},{torusz(\X,\x,\R,\r)});
 \draw[color=red,thick,-latex] 
  plot[smooth,variable=\x,domain={vcrit1(\X,\tdplotmaintheta)}:40,samples=71]   
 ({torusx(\X,\x,\R,\r)},{torusy(\X,\x,\R,\r)},{torusz(\X,\x,\R,\r)});
 }
 % neck
 \draw[color=blue,thick,-latex] plot[smooth,variable=\x,domain=00:360,samples=71]   
 ({torusx(\x,90,\R,\r)},
 {torusy(\x,90,\R,\r)},
 {torusz(\x,90,\R,\r)}); 
 \draw [color=blue,thick](4.3,1.0) node {$\mathbf{b}$};
\draw [color=red,thick](1.0,-3.0) node {$\mathbf{a}$};
\draw [thick](2.6,-1.3) node {$x$};
\draw [thick](2.1,-1.0) node {$\bullet$};
\end{tikzpicture}

%% file: alpha-gamma.tex
\begin{figure}[t]
\centering
%Tikz pictures
%%%% alpha 1
\begin{tikzpicture}[scale=2]
\draw (7.9,-0.4) node {$\alpha$};
\draw [thick,red] (7.4,-0.4) ellipse (0.4 and .2);
\draw [thick,red] (7.3,-0.4) ellipse (0.3 and .15);
\draw [thick,red] (7.2,-0.4) ellipse (0.2 and .1);
\draw [thick,red] (7.1,-0.4) ellipse (0.1 and .05);
\draw [thick](7,-0.4) node {$\bullet$};
\draw [thick](6.9,-0.4) node {$P_{B}$};
%\end{tikzpicture}
%%%
%%%%%% alpha 2
%%%\begin{tikzpicture}[scale=3,tdplot_main_coords]
%\begin{tikzpicture}[scale=2]
%\fill[white] (7.5,0) ellipse (1 and .75);
\draw [thick](7.5,0) ellipse (1 and .75);
% hole
\begin{scope}
  \clip (7.5,-.9) ellipse (1 and 1.25);
  \draw[thick](7.5,1.1) ellipse (1 and 1.25);
  \clip (7.5,1.1) ellipse (1 and 1.25);
  \draw [thick](7.5,-1.1) ellipse (1 and 1.25);
  %\fill[thick,white] (7.5,-1.1) ellipse (1 and 1.25);
\end{scope}
%\draw (7.942,-0.555) node[scale=0.8,rotate=-85] {$<$};
\draw [thick,red] (7.4,-0.4) ellipse (0.4 and .2);
\draw [thick](7.9,-0.4) node {$\alpha$};
\draw [thick](7,-0.4) node {$\bullet$};
\draw [thick](6.9,-0.4) node {$P_{B}$};
\end{tikzpicture}
\vspace{-2cm}
%%%% lambda
\begin{tikzpicture}[scale=2]
%\fill[white,thick] (7.5,0) ellipse (1 and .75);
\draw [thick](7.5,0) ellipse (1 and .75);
% hole
\begin{scope}
  \clip (7.5,-.9) ellipse (1 and 1.25);
  \draw[thick](7.5,1.1) ellipse (1 and 1.25);
  \clip (7.5,1.1) ellipse (1 and 1.25);
  \draw[thick] (7.5,-1.1) ellipse (1 and 1.25);
  %\fill[thick,white] (7.5,-1.1) ellipse (1 and 1.25);
\end{scope}
%\draw (7.942,-0.555) node[scale=0.8,rotate=-85] {$<$};
\draw [thick,blue] (7.5,0) ellipse (0.8 and .47);
\draw [thick,blue] (7.4,0) ellipse (0.75 and .44);
\draw [thick](7.5,.47) node[scale=0.8] {$$} node[above] {$\lambda$};
\draw [thick](7.1,-0.4) node {$\bullet$};
\draw [thick](7.1,-0.5) node {$P_{B}$};
%\draw (7.9,-0.4) node {$\alpha$};
%\draw [thick](7,-0.4) node {$\bullet$};
%\draw (7.4,-0.4) ellipse (0.4 and .2);
\end{tikzpicture}
%
%%%%% mu
\includegraphics [width=4.5cm]{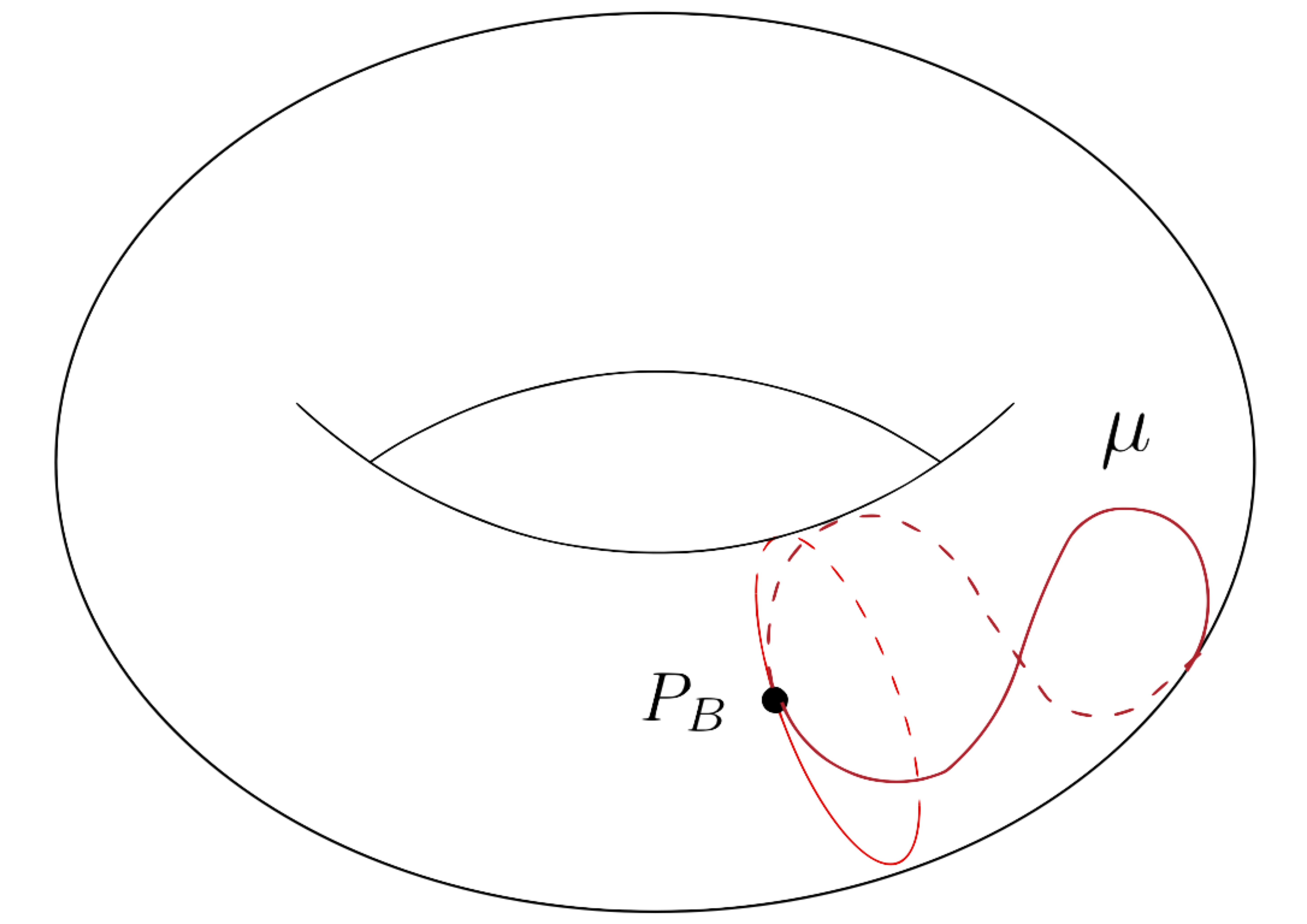} 
%%%% gamma
%------------------------ 3D torus with a complex path
\begin{tikzpicture}[scale=0.4,tdplot_main_coords]
\draw [thick,red](-2.1, -1.25) arc (180:90:0.4cm);
\draw [thick,dashed,rotate=-30, red](1.5,-5) arc (190:300:-2cm);
\draw [thick, rotate=280,red](.6, 2.5) arc (300:180:-1.9cm);
\draw [thick, dashed, rotate=-5,red](1.5, -.5) arc (300:180:-1.9cm);

\draw [thick, rotate=-45,red](2.05, 6.3) arc (300:180:-1.74cm);
%\draw[thick,fill] (3.7,1.1) circle (0.06cm);
\draw[thick,fill] (-2.1,-1.2) circle (0.06cm);
\draw [thick](-2.1,-2.2) node {$P_{B}$};
\draw [thick,fill](4.8,1.1) node{$\gamma$};
\draw[thick,blue] (3.7,1.5) arc (-10:240:3.9cm and 2 cm);
\draw [thick,red](0.2, -7) arc (250:170:0.4cm);
\draw [thick,red](4.33, -4.5) arc (250:170:0.4cm);
\draw [thick,red](1.42, -1.1) arc (180:60:0.3cm);
%
% \draw (-2.8,-1.8) node {$\bullet$};
%\draw (4.5,1.8) node {$|v'\rangle$};
%\draw [step=0.5cm](-6,-8.5) grid (5,10);
%\end{tikzpicture}
%torus
%\begin{tikzpicture}[scale=1,tdplot_main_coords]
\pgfmathsetmacro{\R}{4}
\pgfmathsetmacro{\r}{1.5}
 \draw[thick,,even odd rule,fill opacity=0.2] plot[variable=\x,domain=0:360,smooth,samples=71]
 ({torusx(\x,vcrit1(\x,\tdplotmaintheta),\R,\r)},
 {torusy(\x,vcrit1(\x,\tdplotmaintheta),\R,\r)},
 {torusz(\x,vcrit1(\x,\tdplotmaintheta),\R,\r)}) 
 plot[variable=\x,
 domain={-180+umax(\tdplotmaintheta,\R,\r)}:{-umax(\tdplotmaintheta,\R,\r)},smooth,samples=51]
 ({torusx(\x,vcrit2(\x,\tdplotmaintheta),\R,\r)},
 {torusy(\x,vcrit2(\x,\tdplotmaintheta),\R,\r)},
 {torusz(\x,vcrit2(\x,\tdplotmaintheta),\R,\r)})
 plot[variable=\x,
 domain={umax(\tdplotmaintheta,\R,\r)}:{180-umax(\tdplotmaintheta,\R,\r)},smooth,samples=51]
 ({torusx(\x,vcrit2(\x,\tdplotmaintheta),\R,\r)},
 {torusy(\x,vcrit2(\x,\tdplotmaintheta),\R,\r)},
 {torusz(\x,vcrit2(\x,\tdplotmaintheta),\R,\r)});
 
 % holes
 \draw[thick] plot[variable=\x,
 domain={-180+umax(\tdplotmaintheta,\R,\r)/2}:{-umax(\tdplotmaintheta,\R,\r)/2},smooth,samples=51]
 ({torusx(\x,vcrit2(\x,\tdplotmaintheta),\R,\r)},
 {torusy(\x,vcrit2(\x,\tdplotmaintheta),\R,\r)},
 {torusz(\x,vcrit2(\x,\tdplotmaintheta),\R,\r)});
 \end{tikzpicture}
\caption{ A class of behaviors over a torus $\alpha$ close paths, $\lambda$ path around the neck,  $\mu$ path around the belt,  $\gamma$ complex path} \label{fig:paths}
\end{figure}
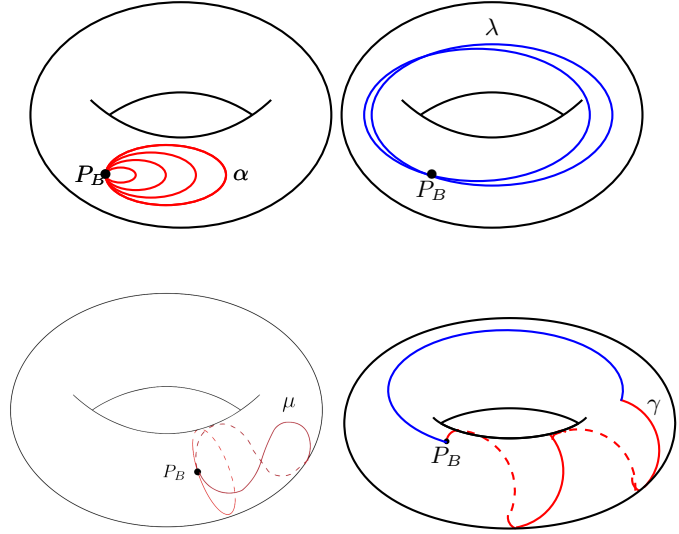